\documentclass{aa}
\usepackage{graphicx}
\usepackage{txfonts}
\begin{document}
   \title{A long look at the BALQSO LBQS~2212-1759 with XMM-Newton}

   \titlerunning{A long look at LBQS~2212-1759 with XMM}

   \author{J. Clavel,
          \inst{1}
          N. Schartel\inst{2} \and L. Tomas\inst{2}
          }

   \offprints{J. Clavel}

   \institute{Research \& Scientific Support Department, 
              ESTEC, SCI-SA, Postbus 299
              2200 AG - Noordwijk, The Netherlands\\
      \email{Jean.Clavel@esa.int}
      \and
       XMM-Newton Science Operations Centre,
       ESAC, Apartado 50727, 28080, Madrid, Spain\\
       \email{nscharte@xmm.vilspa.esa.es, ltomas@xmm.vilspa.esa.es}
       \thanks{This work is based on observations obtained with XMM-Newton,
       an ESA science mission with instruments and contributions directly 
       funded by ESA Member States and the USA (NASA).
	     }
             }

   \date{Received: 7 September 2004 / Accepted: 9 September 2005}

   \abstract{Very long (172~ks effective exposure time) observations of the 
   BALQSO \object{LBQS~2212-1759} with {\em XMM-Newton\/} yield a 
   stringent upper-limit on its 0.2--10~keV (rest- frame 0.64--32.2~keV) flux,
   ${\rm F_{0.64-32.2} \leq 6\times10^{-17}~erg\,cm^{-2}\,s^{-1}}$, while 
   simultaneous UV and optical observations reveal a rather blue spectrum 
   extending to 650~\AA\/ in the source rest frame. These results are
   used to set a tight upper-limit on its optical to X-ray spectral index 
   ${\rm \alpha_{ox} \leq -2.56}$. Given the HI-BAL nature of 
   \object{LBQS 212-1759}, its X-ray weakness is most likely due to 
   intrinsic absorption. If this is the case, and assuming that the intrinsic
   ${\rm \alpha_{ox}}$ of \object{LBQS 2212-1759} is -1.63 -- a value 
   appropriate for a radio-quiet quasar of this luminosity -- one can set 
   a lower limit on the X-ray absorbing column 
   ${\rm N_{H} \geq 3.4\,10^{25} cm^{-2}}$. Such a large column has a Thomson 
   optical depth to electron scattering ${\rm \tau_{Th} \geq 23}$, sufficient
   to extinguish the optical and UV emission. The problem only gets worse if 
   the gas is neutral since the opacity in the Lyman continuum becomes 
   extremely large, ${\rm \tau_{Ly}\,\geq\,2\times 10^{8}}$, conflicting with 
   the source detection below 912~\AA\/. This apparent contradiction 
   probably means that our lines-of-sight to the X-ray and to the UV emitting
   regions are different, such that the gas covers completely the 
   compact X-ray source but only partially the more extended source of
   ultraviolet photons. An extended ($\simeq 1\arcmin$) X-ray source is 
   detected $\sim 2\arcmin$ to the south-east of the QSO. Given its thermal 
   spectrum and temperature (${\rm 1.5\,\leq\,T\,\leq\,3.0 keV}$), it is 
   probably a foreground ($0.29 \leq z \leq 0.46$) cluster of galaxies.
   
   }

   \maketitle
%

\section{Introduction}

   Broad Absorption Line (BAL) quasars are characterized by broad and
   blueshifted absorption troughs in their spectrum, from resonance 
   transitions such as CIV$\lambda$1550, Ly$\alpha\lambda$1216, 
   NV$\lambda$1240, indicating the presence of a high velocity (up to 
   50,000~${\rm km\,s^{-1}}$) outflow along the line-of-sight (LOS) to the 
   nucleus. Taking into account selection biases, BALQSOs represent 
   22$\pm$4\,\% of the radio-quiet quasar population (Hewett and Foltz 
   \cite{hewett}). The fraction of BALQSOs that are radio-loud is
   approximately the same as that of non-BAL quasars, but there appears 
   to be a deficit of broad absorption line objects at large radio 
   luminosities (Menou et al. \cite{menou}; Becker et al. \cite{becker}). 
   Because of the overall similarity of their continuum and emission line 
   properties with those of non-absorbed quasars, it is sometime thought that
   BALQSOs are ``normal'' quasars seen at a specific viewing angle such that 
   our LOS intercepts a nuclear wind (Weyman et al.\cite{weyman}). The wind
   possibly originates from the accretion disk and is driven out radially 
   by radiation pressure (Murray et al. \cite{murray}).
   A hydrodynamic model for the BAL wind was developed by Proga 
   et al. (\cite{proga}). In the empirical scenario proposed 
   by Elvis (\cite{elvis}), the wind arises vertically from a
   narrow range of disk radii and bends outward to a cone angle
   of ~60$^o$ with a divergence angle of ~6 $^o$. In this type of models, it 
   is the solid angle covered by the outflow that determines the fraction
   of BAL quasars. 
   An alternative class of models speculates that the BAL phenomenon 
   represents an early ``cocoon'' phase in the evolution of a QSO (e.g. Briggs,
   Turnshek and Wolfe \cite{briggs}). Although his results are based on 
   a small sample that contains only 4 BALQSOs, Boroson \cite{boroson} 
   brought some credibility to this idea by showing that BALQSOs occupy a 
   specific location in the quasar parameter space, characterized by large 
   accretion rates and luminosities, close to the Eddington limit. 
   The evolution scenario is also supported by the large fraction
   of BALQSOs found in a spectroscopic follow-up to the VLA FIRST
   survey -- 29 radio-selected BALQSOs (Becker et al. \cite{becker}) --
   since the properties of the sample appear inconsistent with simple
   unified models.

   BALQSOs are invariably X-ray weak or silent (Green et al. \cite{green95}; 
   Green \& Mathur \cite{green96};
   Gallagher et al. \cite{gallagher99}), suggesting the presence of very large 
   absorbing columns, ${\rm N_{H} \geq 10^{23}\,cm^{-2}}$, 2-3 orders of 
   magnitudes larger than those inferred from UV absorption line studies. This 
   discrepancy led to the conclusion that the bulk of the absorbing gas is 
   highly ionised and thus mostly transparent in the ultraviolet while still 
   providing large X-ray opacities. However,
   it was subsequently realized that the column densities derived from curve of
   growth analysis of absorption lines may be severely underestimated. High
   resolution and high signal-to-noise ratio UV spectra show that the lines 
   are saturated despite the existence of residual flux at their bottom (e.g.
   Arav et al. \cite{arav}; Wang et al. \cite{wang}). The residual flux
   may be due to partial covering of the continuum source or to the scattering 
   of part of its emission back into our LOS, as indicated by the higher
   degree of polarisation of BALQSOs as compared to non-BAL quasars
   (e.g. Schmidt and Hines \cite{schmidt}; Ogle et al. \cite{ogle}).
   

   Here we present very sensitive observations of the z = 2.217 BALQSO 
   LBQS~2212-1759 (Morris et al.\cite{morris}) performed with 
   {\em XMM-Newton\/}. This quasar was selected because of its optical 
   brightness (${\rm m_{B}\,=\,17.94}$) and tentative 3-sigma detection in 
   the soft-X-ray band with ROSAT (Green et al.\cite{green95}).
   LBQS~2212-1759 displays two CIV$\lambda$1548 absorption troughs 
   blue-shifted respectively by $\simeq$ -6,300 and -4,000~${\rm km s^{-1}}$  
   with respect to its systemic velocity (Korista et al. \cite{korista}).
    


\section{Observations and data reduction}
   LBQS~2212-1759 was observed twice with {\em XMM-Newton\/} (Jansen et al.
   \cite{jansen}). The first observation took place from 22h15m (U.T.) on 
   November 17, 2000 to 17h44m on November 18, under conditions of extremely 
   low particle radiation background. The quasar was observed again one year 
   later, from 22h17m on November 17, 2001 to 04h53m on November 19. On each 
   occasion, data were collected simultaneously with all X-ray instruments on 
   board {\em XMM-Newton}: the {\em EPIC-pn\/} (Str\"{u}der et al. 
   \cite{struder}), the two {\em EPIC-MOS\/} (Turner et al.\cite{turner}), 
   as well as the two {\em RGS\/} spectrographs (den Herder et al. 
   \cite{den herder}{\bf)}. As expected, the last 
   did not yield any useful information and the corresponding data are not 
   discussed further in the remainder of this paper. A log of the X-ray 
   observations is provided in Table~\ref{Log}, where we give the 
   {\em XMM-Newton\/} observation identifier in column~1, the start date 
   of the observation in column~2 and the exposure times with the three 
   {\em EPIC pn\/} and {\em EPIC MOS\/} instruments in column 3 \& 4,
   respectively. Note the very long cumulative exposure time of the X-ray 
   observations. In the case of the most sensitive {\em EPIC-pn\/} instrument, 
   it reaches 221,352 s. Even after data screening 
   the cumulative useful integration time of the {\em EPIC pn\/} data 
   remains 172,628 s. 

      \begin{table}
   	 \caption[]{Details of the XMM-Newton X-ray observations}
   	    \label{Log}
   	    \centering
	 \begin{tabular}{c c c c} 
         \hline\hline
	Obs ID     & Start Date & \multicolumn{2}{c}{exposure times (s)}\\   
        	   &            & EPIC pn & EPIC MOS 1\&2     \\
         \hline
	0106660101 & 2000-11-17 &  55,718 &  2$\times$57,824  \\
	0106660201 & 2000-11-18 &  50,618 &  2$\times$52,724  \\
        0106660401 & 2001-11-18 &     0   &  2$\times$33,050  \\
	0106660501 & 2001-11-17 &   8,208 &  2$\times$10,822  \\
	0106660601 & 2001-11-17 & 106,808 &  2$\times$109,422 \\
	\hline                  
	\end{tabular}
	\end{table}

      \begin{table}
   	 \caption[]{$3-\sigma$ Upper limits to the X-ray flux of 
	 \object{LBQS 2212-1759} derived from the EPIC-pn observations}
   	    \label{xlim}
   	    \centering
	 \begin{tabular}{c c c} 
         \hline\hline
energy band  &  Count rate               & Flux                                    \\
  (keV)      & (${\rm 10^{-5}\,s^{-1}}$) & (${\rm 10^{-16}\,erg\,cm^{-2}\,s^{-1}}$)\\
         \hline
0.2--0.5     &  $\leq 9.03$              & $\leq 1.69$                             \\
0.5--2.0     &  $\leq 9.25$              & $\leq 2.74$                             \\
2.0--10.0    &  $\leq 8.65$              & $\leq 14.2$                             \\
5.0--10.0    &  $\leq 6.16$              & $\leq 20.1$                             \\
0.2--10.0    &  $\leq 1.48$              & $\leq 0.64$                             \\
	\hline                  
	\end{tabular}
	\end{table}

   The X-ray {\em pn\/} and {\em MOS\/} data were reduced and analyzed in a 
   standard fashion using the SAS v5.3.
   The pipe-line products of observation 0106660601 provide 44 
   positional coincidences between sources of the USNO-A2.0 Catalogue
   (Monet et al. \cite{monet}) and X-ray sources in the field of 
   LBQS~2212-1759. Of these 44 coincidences, 35 X-ray sources have only 
   one, three X-ray sources have two and one X-ray source has three optical 
   counterparts. Restricting to the 35 X-ray sources with a unique optical 
   identification, we infer a mean offset of 2.2\arcsec between the X-ray 
   position and the optical source coordinates. Note that out of these 35
   X-ray sources with a unique identification, 25 (i.e. 63\%) lie within 
   2\arcsec of their optical counterpart. The BAL quasar was not detected 
   in either of the X-ray instruments. The nearest detected X-ray point 
   source is $\sim0.5$\arcmin away from the nominal position of 
   \object{LBQS~2212-1759} (R.A. = 22:15:31.6; Dec. = 17:44:06 - J2000)

   From the {\em r.m.s.\/} 
   background count fluctuations in a $9\times9$-pixel cell centered on the 
   expected source position, we computed $3~\sigma$ upper limits to the count
   rate in various energy bands.
   The HEW of the EPIC-pn point-spread function is 14\arcsec and one 
   CCD pixel projects onto an area of 4.1$\times$4.1\arcsec on the sky.
   From the XMM-Newton observation of  Q~0056-363, we determined the ratios
   between the total count rate of a faint point-source and the
   count rates measured in a $9\times9$-pixel cell centered on the source
   for each of the energy bands. These ratio were then applied to the 
   cell upper limits to derive effective upper limits to the source
   count rate. The results are listed in table \ref{xlim}.
   These count rates were converted into flux
   upper limits using the {\em PIMMS\/} software available on-line at the 
   HEASARC web-site. The results are given in Table~\ref{xlim} were we only
   list the results from the {\em EPIC pn\/} data, since the less sensitive
   {\em EPIC MOS\/} detectors yield consistent but significantly higher 
   and therefore less constraining limits. 
   
   In an attempt to understand why \object{LBQS~2212-1759} was
   marginally detected by Green et al. \cite{green95}, we checked their
   original ROSAT image. There are definitely no excess counts at the 
   centre of their 3\arcmin radius extraction circle, clearly ruling out
   the presence of a point-source. However, the merged EPIC data reveal the
   existence of a weak {\em extended\/} source, centered at R.A. = 22:15:37
   and Dec=-17:45:35 and whose radius is 1.0\arcmin. This source is most
   probably a foreground cluster of galaxies since its spectrum is well 
   described by a Mekal spectrum with temperature in the range 1.5-3 keV 
   and a redshift between 0.29 and 0.46. Whatever its origin, this source
   clearly lies within the ROSAT extraction region and is likely the origin 
   of the false detection of \object{LBQS~2212-1759} by Green et al. 
   \cite{green95}. The extended source was too weak to appear in the ROSAT All 
   Sky Survey catalogue (1RXS) and 
   could not therefore be taken into account by Green et al.\cite{green95}.
   Note that there are no EPIC point-sources within the 10-20\arcmin annulus 
   which these authors used to measure the background in the ROSAT image.
   
   In parallel to the X-ray observations, a series of optical and ultraviolet 
   broad-band filter images of the QSO field were obtained with the Optical
   Monitor telescope ({\em OM\/}; Mason et al. \cite{mason}) on board 
   {\em XMM-Newton\/}. The {\em OM\/} data were reprocessed 
   with the SAS version 6.0 using the script {\em omichain\/}. For each 
   broad-band filter image, the corrected net count rate of the QSO was 
   read-off directly from the SWSRLI output files, which lists all sources 
   automatically detected by the SAS software, together with their count
   rate, statistical significance, measured coordinates, the associated 
   errors and various data quality indicators. We used the close-by 
   ${\rm 13^{th}}$ magnitude star {\em S3211320188\/} from the HST guide-star 
   catalog to correct for small ($\leq 4\arcsec$) residual astrometric distortions 
   in the {\em OM\/} coordinate system. After correction, the QSO coordinates
   as measured with {\em OM\/} agree to better than $1\arcsec$ with {\em NED\/}
   catalog coordinates. The count rates were converted 
   into fluxes following the recipe provided on the SAS web page at URL 
   xmm.vilspa.esa.es/sas/documentation/watchout/uvflux.html. The final fluxes
   are listed in Table~\ref{OMFlux}, which provides: the observation identifier
   in column~1, the OM exposure number in column~2, the date and U.T. time 
   of the  start of the exposure expressed as a fractional day of 2000 in 
   column~3, the filter identifier in column~4 and the flux with its associated 
   statistical {\em r.m.s.\/} error in column~5. Early observations with
   the less sensitive UVW2 filter had exposure times that were too short and 
   did not yield statistically significant detections. In such cases, 
   $3-\sigma$ upper limits are listed in Table~\ref{OMFlux}.
   
      \begin{table*}
   	 \caption[]{Optical \& UV fluxes measured though the various OM filters}
   	    \label{OMFlux}
   	    \centering
	 \begin{tabular}{c c c c c} 
         \hline\hline

Obs ID    & Exposure \# & Start date    & Filter & Flux          \\
          &             & (2000 day \#) &        & (${\rm 10^{-16}\,erg\,cm^{-2}\,s^{-1}\,\AA^{-1}}$) \\
         \hline
106660101 &  10  &   321.78944   &  V	 &    $2.79\pm0.33$   \\
106660101 & 401  &   321.80242   &  V	 &    $3.32\pm0.33$   \\
106660101 & 402  &   321.81542   &  V	 &    $2.72\pm0.33$   \\
106660101 & 403  &   321.82841   &  V	 &    $2.80\pm0.32$   \\
106660101 & 404  &   321.84140   &  V	 &    $3.00\pm0.33$   \\
106660201 &   6  &   322.54738   &  V	 &    $2.77\pm0.30$   \\
106660201 & 401  &   322.56269   &  V	 &    $2.42\pm0.30$   \\
106660201 & 402  &   322.57797   &  V	 &    $2.41\pm0.30$   \\
106660201 & 403  &   322.59329   &  V	 &    $2.47\pm0.29$   \\
106660101 &   8  &   321.91927   &  B	 &    $4.33\pm0.22$   \\
106660101 & 409  &   321.93226   &  B	 &    $4.36\pm0.22$   \\
106660101 & 410  &   321.94523   &  B	 &    $4.42\pm0.22$   \\
106660101 & 411  &   321.95822   &  B	 &    $4.25\pm0.21$   \\
106660101 & 412  &   321.97122   &  B	 &    $3.95\pm0.22$   \\
106660201 &   9  &   322.70034   &  B	 &    $4.23\pm0.20$   \\
106660201 & 409  &   322.71564   &  B	 &    $4.17\pm0.20$   \\
106660201 & 410  &   322.73094   &  B	 &    $4.12\pm0.20$   \\
106660201 & 411  &   322.74624   &  B	 &    $4.35\pm0.20$   \\
106660201 & 412  &   322.76154   &  B	 &    $4.02\pm0.20$   \\
106660101 &   7  &   321.85435   &  U	 &    $3.84\pm0.20$   \\
106660101 & 405  &   321.86733   &  U	 &    $4.13\pm0.20$   \\
106660101 & 406  &   321.88032   &  U	 &    $4.12\pm0.20$   \\
106660101 & 407  &   321.89332   &  U	 &    $4.11\pm0.20$   \\
106660101 & 408  &   321.90630   &  U	 &    $3.59\pm0.20$   \\
106660201 &   8  &   322.62387   &  U	 &    $4.00\pm0.18$   \\
106660201 & 405  &   322.63916   &  U	 &    $3.83\pm0.18$   \\
106660201 & 406  &   322.65446   &  U	 &    $3.84\pm0.18$   \\
106660201 & 407  &   322.66976   &  U	 &    $4.24\pm0.18$   \\
106660201 & 408  &   322.68506   &  U	 &    $3.73\pm0.18$   \\
106660101 &   9  &   321.98869   &  UVW1 &    $3.85\pm0.20$   \\
106660101 & 413  &   322.01070   &  UVW1 &    $3.70\pm0.20$   \\
106660101 & 414  &   322.05354   &  UVW1 &    $3.60\pm0.20$   \\
106660101 & 415  &   322.07557   &  UVW1 &    $3.57\pm0.19$   \\
106660101 & 416  &   322.09757   &  UVW1 &    $3.73\pm0.20$   \\
106660201 &  10  &   322.78260   &  UVW1 &    $3.65\pm0.18$   \\
106660201 & 413  &   322.80948   &  UVW1 &    $3.68\pm0.18$   \\
106660201 & 414  &   322.83636   &  UVW1 &    $3.59\pm0.18$   \\
106660201 & 415  &   322.86323   &  UVW1 &    $3.60\pm0.18$   \\
106660201 & 416  &   322.89011   &  UVW1 &    $3.68\pm0.18$   \\
106660501 &  6   &   686.10035   &  UVW1 &    $3.39\pm0.17$   \\
106660601 & 11   &   687.44819   &  UVW1 &    $3.48\pm0.11$   \\
106660101 & 10   &   322.11957   &  UVM2 &    $3.28\pm0.39$   \\
106660101 & 418  &   322.16359   &  UVM2 &    $2.78\pm0.39$   \\
106660101 & 419  &   322.18561   &  UVM2 &    $2.97\pm0.39$   \\
106660101 & 420  &   322.20762   &  UVM2 &    $2.99\pm0.39$   \\
106660401 &   9  &   685.93929   &  UVM2 &    $2.41\pm0.26$   \\
106660601 &  13  &   687.65182   &  UVM2 &    $2.50\pm0.24$   \\
106660101 &  11  &   322.23667   &  UVW2 &    $<6.60$         \\
106660101 & 422  &   322.30894   &  UVW2 &    $<7.19$         \\
106660101 & 423  &   322.34508   &  UVW2 &    $<6.56$         \\
106660101 & 424  &   322.38120   &  UVW2 &    $3.48\pm0.70$   \\
106660401 &  10  &   686.02503   &  UVW2 &    $1.33\pm0.58$   \\
106660501 &   8  &   686.15948   &  UVW2 &    $<8.16$         \\
106660601 & 14   &   687.74334   &  UVW2 &    $2.20\pm0.58$   \\
106660601 & 15   &   687.83490   & UVW2  &    $2.15\pm0.58$   \\
 \hline 				        	   
 \end{tabular}
 \end{table*}
     
    A $\chi^{2}$ test shows that the flux of \object{LBQS~2212-1759}  
    remained constant within the measurement uncertainties in all 6 filters. 
    The reduced chi-squares (d.o.f.) corresponding to the hypothesis of a 
    constant flux are $\chi^{2}_{\nu} \,=\,$ 0.88 (8), 0.53(9), 1.19 (9),
    0.52 (10) and 1.20 (6) for the V, B, U, UVW1 and UVM2 filters, 
    respectively.
    We therefore averaged the results 
    from individual exposures and computed the weighted mean flux in each 
    filter and the error on the mean. The near-IR {\em J, H} and {\em K\/} 
    fluxes of 
    \object{LBQS~2212-1759} were retrieved from the {\em 2MASS\/} catalog 
    (Kleinmann \cite{kleinmann}; Barkhouse and Hall \cite{barkhouse}). All 
    fluxes were finally corrected for foreground galactic extinction 
    (${\rm E_{b-v} = 0.026}$; Schlegel et al. \cite{schlegel}; Cardelli et 
    al. \cite{cardelli}). The results are given in Table~\ref{MeanFlux},
    where we list the origin of the data in column~1, the effective wavelength
    and band-pass of the filter in the observer's frame in column~2 and 
    column~3, respectively, the effective wavelength and band-pass of the filter
    in the quasar rest-frame in column~4 and column~5, respectively and the 
    averaged de-reddened flux in column~6.

      \begin{table*}
   	 \caption[]{The flux of \object{LBQS~2212-1759} as a function of
	 wavelength, from the near-IR to the EUV. The values in the
	 optical \&  UV have been obtained by averaging fluxes from individual
	 OM exposures. All fluxes have been corrected for foreground 
	 galactic reddening}
   	    \label{MeanFlux}
   	    \centering
	 \begin{tabular}{c c c c c c} 
         \hline\hline
Instrument  & ${\rm \lambda_{obs}}$ & ${\rm \Delta\lambda_{obs}}$ & ${\rm \lambda_{rest}}$ & ${\rm \Delta\lambda_{rest}}$ & Flux \\
            &  \multicolumn{4}{c}{\it (\AA)} & (${\rm 10^{-16}\,erg\,cm^{-2}\,s^{-1}\,\AA^{-1}}$) \\
         \hline
2MASS K & 21,590               &  2,620                  & 6,711 & 814 & $0.945\pm0.047$    \\  
2MASS H & 16,620               &  2,510                  & 5,166 & 780 & $1.006\pm0.071$    \\ 
2MASS J & 12,350               &  1,620                  & 3,839 & 504 & $0.987\pm0.082$    \\  
OM-V    & 5,430                &  710                    & 1,688 & 221 & $2.957\pm0.040$    \\
OM-B    & 4,500                &  980                    & 1,399 & 305 & $4.673\pm0.018$    \\
OM-U    & 3,440                &  780                    & 1,069 & 242 & $4.478\pm0.026$    \\
OM-UVW1 & 2,910                &  710                    & 905   & 221 & $4.186\pm0.013$    \\
OM-UVM2 & 2,310                &  460                    & 718   & 143 & $3.447\pm0.082$    \\
OM-UVW2 & 2,120                &  460                    & 659   & 143 & $2.868\pm0.379$    \\
 \hline 				        	   
 \end{tabular}
 \end{table*}

\begin{figure*}
  \centering
  \includegraphics[width=17cm]{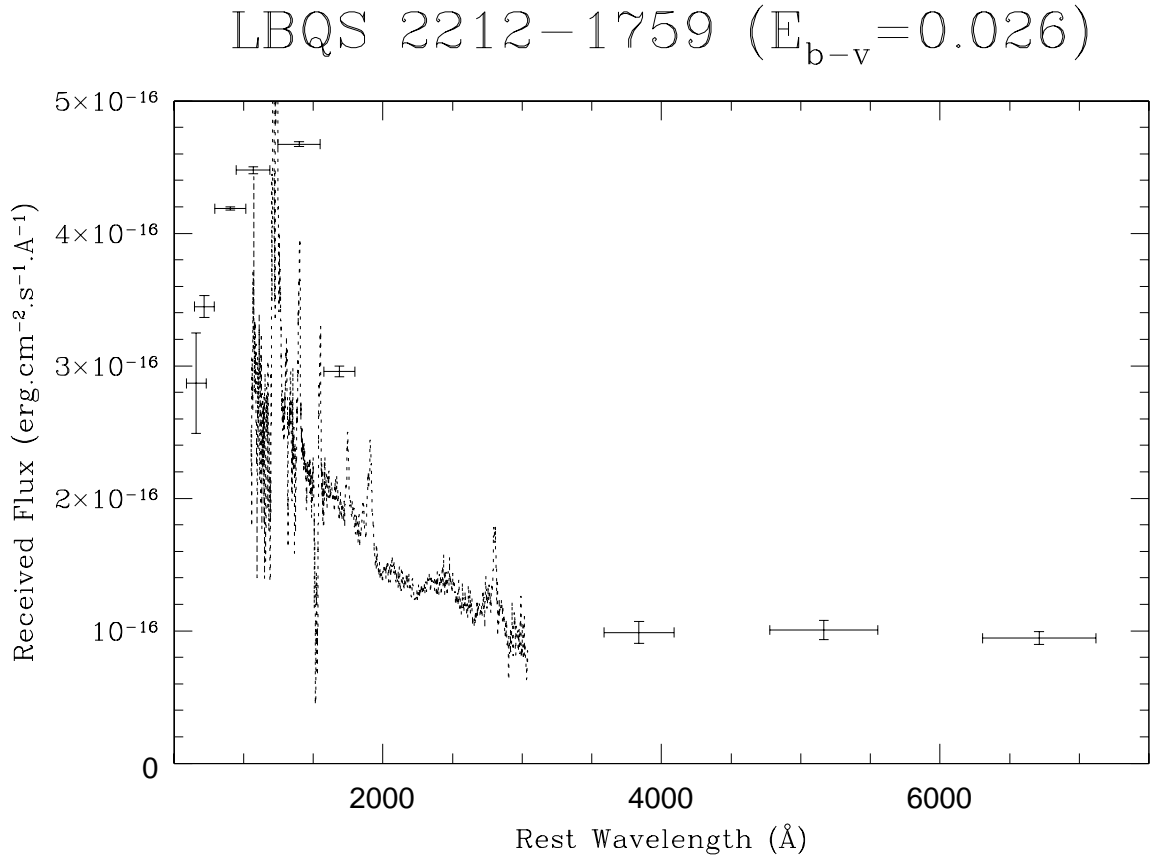}
  \caption{The flux distribution of LBQS~2212-1759 from 6,711~\AA\/ to
   659~\AA\/ (rest wavelengths). All data-points are from the present study,
   except for the 3 longest wavelengths ones which were retrieved from the
   2MASS catalogue. The 1992 spectrum from Korista (1993) is also shown
   for comparison. All flux values are in the observer's frame and have been 
   corrected for foreground galactic extinction (${\rm E_{b-v} = 0.026}$).}
  \label{UVOpt_Spec}
  \end{figure*}

   The rest-frame optical to EUV energy distribution of LBQS~2212-1759, 
   is shown in Figure~\ref{UVOpt_Spec}. A spectrum of 
   \object{LBQS 2212-1759} obtained by K. Korista in 1992 (Korista 
   et al.\cite{korista}) and kindly provided to us by the author is also 
   displayed for comparison. Note that while the flux increased by 46~\% in 
   the OM-V band during the 8.5 years interval between the two observations,
   the spectral shape remained very similar.
 
  \begin{figure*}
  \centering
  \includegraphics[width=17cm]{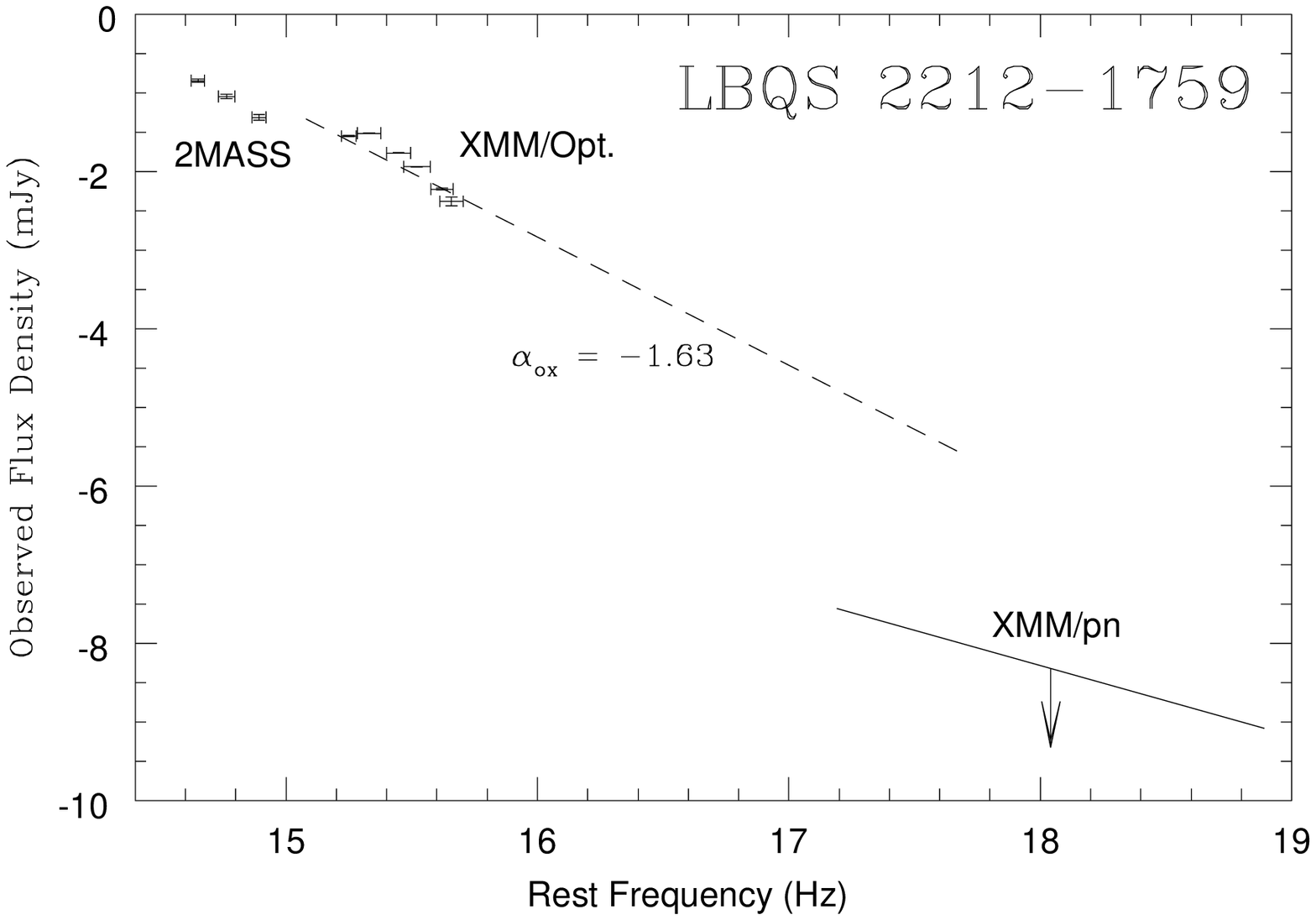}
  \caption{The overall optical-to-X-ray spectral energy distribution of 
   LBQS~2212-1759. All flux values are in the observer's frame and have been 
   corrected for foreground galactic extinction (${\rm E_{b-v} = 0.026}$).
   The upper limit to the 0.2--10 keV (observer's frame) flux of table 
   \ref{xlim} is shown as a spectrum of photon index $\Gamma = -1.9$.
   For comparison, a fiducial power-law of index $\alpha_{ox} = -1.63$ 
   and extending from 2500~\AA\/ to 2 keV (rest frame) is shown as a dashed 
   line. The LBQS~2212-1759 flux upper limit is 263 times lower than the 
   flux predicted by extrapolating the ${\rm \alpha_{ox}}$ power-law to 2 keV}
  \label{SED}
  \end{figure*}

  \section{Implications and discussion}
   The UV-optical SED of \object{LBQS 2212-1759} is typical of a high 
   redshift QSO; at wavelengths longer than that of H-Ly$\alpha\lambda$1215, 
   it displays a blue spectrum giving rise to an excess emission, the
   so-called ``big blue bump'', usually attributed to the thermal emission of
   an accretion disk. A power-law (${\rm F_{\nu} \propto \nu^{\alpha}}$) fit 
   to the 5 data-points with ${\rm \lambda_{rest} \geq 1215\,\AA\/}$  yields a 
   spectral index $\alpha = -0.96\pm0.18$. Such an index is within 
   the range of optical-UV slopes observed in the general quasar population.  
   As noted by several authors (see e.g. Elvis et al. \cite{elvis94}), the 
   spread in indices is fairly large. In the Francis et al. (\cite{francis}) 
   sample of 688 LBQS quasars for instance, indices vary from -1.5 to +1. 
   The 1050--2200 \AA\/ spectral index of the average composite quasar of 
   Zheng et al. (\cite{zheng}) is -0.86$\pm$0.01, while the mean 
   1285--5100~\AA\/ spectral index of the radio-quiet QSO sample of Kuhn 
   et al. (\cite{kuhn}) is -0.32$\pm$0.28 (1-$\sigma$) and independent
   of redshift. The composite quasar spectrum of Vanden Berk et al. (\cite 
   {vanden berk}), obtained by averaging ~2200 QSO spectra from the Sloan
   Digital Sky Survey, has a 1350--4230~\AA\/ index of -0.44. At wavelengths 
   shorter than 1215\,\AA\/, the \object{LBQS 2212-1759} spectrum steepens
   to a softer spectral index $\alpha = -2.62\pm0.16$. Again, such 
   a steepening is not unusual amongst quasars and has been reported by 
   several authors (e.g. Zheng et al. \cite{zheng}; Kuhn et al. \cite{kuhn};
   Vanden Berk et al. \cite{vanden berk}). 
   In a ``normal'' high redshift quasar, this sharp steepening of the 
   spectrum in the EUV is due to the onset of many intervening Lyman series 
   absorption systems along the line-of-sight, the so-called ``Lyman alpha 
   forest''. In a BALQSO like \object{LBQS 2212-1759}, the steepening can
   also be attributed, at least partly, to resonance absorption within the BAL
   nuclear outflow.

   The ultraviolet spectrum of \object{LBQS 2112-1759} is however difficult 
   to reconcile with the above stringent upper limits on its X-ray flux. 
   The optical-to-X-ray spectral index of a quasar (Zamorani et al. 
   \cite{zamorani}), ${\rm \alpha_{ox}}$, is defined as the spectral index of 
   an hypothetical power-law connecting its flux density at 
   2500~\AA\/ and 2.0~keV in the QSO rest-frame,
   ${\rm \alpha_{ox}\,=\,0.384\,\log{\frac{F_{2\,keV}}{F_{2500}}}}$. 
   In radio-quiet non-BAL quasars, it is observationally confined to a range 
   ${\rm -1.1 \geq \alpha_{ox} \geq -1.9}$ with a weak dependence on the 
   source luminosity (Vignali et al. \cite{vignali}; Strateva et al.
   \cite{strateva}). Using the same cosmological parameters
   as these authors, the monochromatic luminosity of \object{LBQS 2112-1759}
   at 2500~\AA\/ (rest wavelength) is
   ${\rm L_{2500}\,=\,1.7\,10^{31}\,erg\,s^{-1},Hz{-1}}$, which, 
   according to eq. 6 of Strateva et al., predicts 
   ${\rm \alpha_{ox}\,=\,-1.63\pm0.03}$.
   Assuming a canonical photon spectral index $\Gamma\,=\,-1.9$ for the
   0.2--10~keV spectrum of \object{LBQS 2212-11759} (e.g. Laor et al. 
   \cite{laor}), and using the X-ray flux upper limits
   of Table~\ref{xlim}, one can infer an upper limit to the monochromatic
   flux density at a rest energy of 2~keV (${\rm E_{obs} = 0.621~keV}$), 
   ${\rm F_{\nu}(2keV) \leq 1.0\,10^{-8}\,mJy}$. One can derive 
   the flux at ${\rm \lambda_{rest} = 2500\,\AA\/}$ by interpolation
   between the de-reddened fluxes in the J and V bands. Combining the two 
   yields $\alpha_{ox} \leq -2.56$, steeper by $\sim\,1$~dex than the 
   index predicted for a radio-quiet quasar of the same luminosity as
   \object{LBQS 2212-1759}. This is illustrated in Figure~\ref{SED},
   where we plot the overall Spectral Energy Distribution (SED) of 
   \object{LBQS 2212-1759}. The upper limit to its 2~keV flux is 263
   times lower than that predicted by extrapolation of its 2500~\AA\/ flux
   density with a power-law of index ${\rm \alpha_{ox}\,=\,-1.63}$.
   Assuming that the difference is entirely due to intrinsic absorption of the
   X-ray flux, one can infer a lower limit to the required absorbing column,
   ${\rm N_{H} \geq 3.4\,10^{25}\,cm^{-2}}$. Note that this result
   depends only weakly on the value assumed for the X-ray spectral index.
   For instance, using $\Gamma\,=\,-2.5$ instead of -1.9 hardly changes the 
   results to ${\rm N_{H} \geq 2.6\,10^{25}\,cm^{-2}}$.

   If the gas was neutral, an absorbing column as large as or larger than  
   ${\rm 3.4\,10^{25}\,cm^{-2}}$ would create an optical depth at the 
   Hydrogen Lyman limit, ${\rm \tau_{Ly}\,\geq\,2.2\,10^{8}}$, more than 
   sufficient to extinguish all radiation at wavelengths shorter than
   912~\AA\/. This is not the case however, since \object{LBQS 2212-1759} is
   detected to ${\rm \lambda_{rest}\,=\,659\,\AA\/}$. Hence, the X-ray 
   absorbing gas cannot be neutral and cover the UV continuum source. 
   However, even if the gas is fully ionised,
   the Thomson optical depth to electron scattering corresponding to the above 
   column, ${\rm \tau_{th} \geq 23}$, is sufficient to attenuate the flux 
   by a factor $\simeq 10^{10}$ and make \object{LBQS 2212-1759} invisible at
   all wavelengths except in the $\gamma$ ray regime. Another difficulty is
   that, unless the gas is completely free of dust, extinction will wipe-out 
   any emerging UV and optical photon. Even if the dust to gas ratio is 
   100 times lower than the average galactic value (e.g. Gorenstein 
   \cite{gorenstein}), an absorbing column of ${\rm 3.4\,10^{25}\,cm^{-2}}$ 
   would still generate $\sim$150 magnitudes of visual extinction and 
   approximately ten times more in the far ultraviolet.
   
   We are thus left with an inconsistency: on the one hand, \object{LBQS
   2212-1759} is detected with high statistical significance in the UV 
   and EUV range, and on the other it is not detected in the X-rays, with
   upper limits on the 0.2--10~keV flux which, taken at face value, imply 
   column densities sufficient to extinguish its ultraviolet emission as well. 

   In what follows, we briefly explore two possible explanations for this
   apparent contradiction:

   \begin{figure}
   \resizebox{\hsize}{!}{\includegraphics{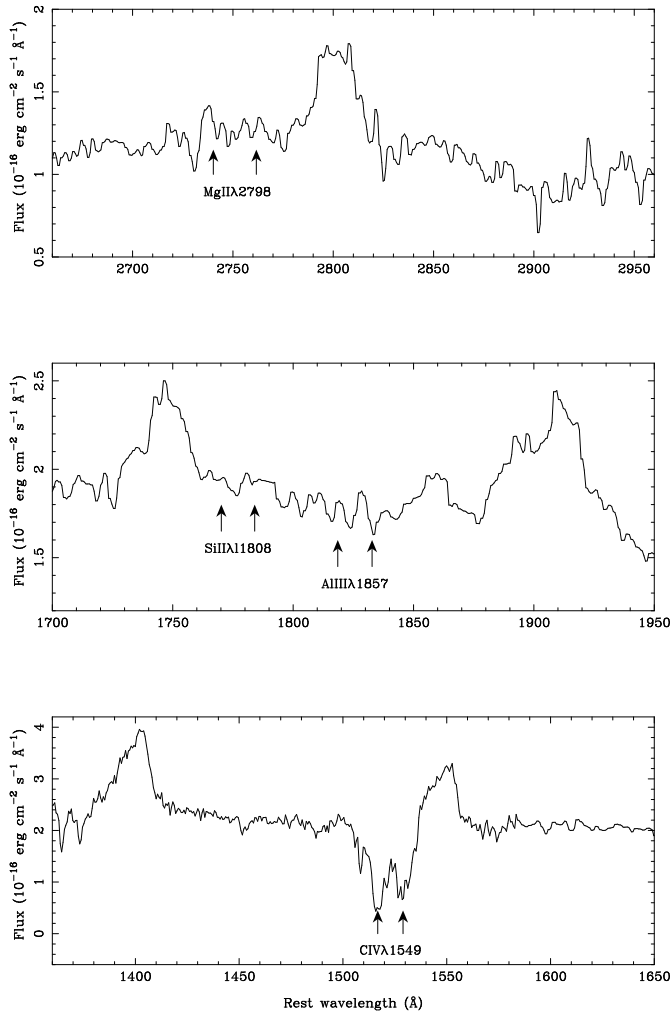}}
   \caption{The spectrum of \object{LBQS 2212-1759} around the MgII$\lambda$2798
   line (upper panel), the AlIII$\lambda$1857 and SiII$\lambda$1808 lines
   (middle), and the CIV$\lambda$1549 line (bottom). The wavelengths at which 
   one expects blue-shifted double absorption troughs are indicated with arrows.
   Note the absence of significant absorption from MgII, SiII and AlIII.} 
   \label{abslines}
   \end{figure}

  \begin{table}
  \caption{Intensities and full width at half maximum (FWHM) of the main 
  emission lines in the spectrum of LBQS~2212-1759; all intensities are in 
  units of ${\rm 10^{-16}\,erg\,cm^{-2}\,s^{-1}}$ and have been corrected 
  for galactic extinction. Their uncertainties are typically 10~\%. The FWHM
  are expressed in ${\rm km\,s^{-1}}$ and have an uncertainty of 
  ${\rm \pm50\,km\,s^{-1}}$}
  \label{Lines}
  \centering
  \begin{tabular}{lcc}
  \hline \hline
  Line ID & Flux & FWHM \\
  \hline
  H-Ly$_{\alpha}\lambda$1216 & 204      & 1610 \\
  NV$\lambda$1240            & 132      & 4780 \\
  SiIV$\lambda$1397          & 50       & 3600 \\
  CIV$\lambda$1549           & 33       & 2790 \\
  HeII$\lambda$1640          & $\leq$ 4 & --   \\
  NIII]$\lambda$1750         & 25       & 2780 \\
  CIII]$\lambda$1909         & 49       & 5150 \\
  MgII$\lambda$2798          & 41       & 2590 \\
  \hline
  \end{tabular}
  \end{table}

   \begin{enumerate}
   \item{{\it \object{LBQS 2212-1759} is genuinely X-ray weak, with an
   intrinsic optical to X-ray spectral index $\alpha_{ox} \leq -2.56$}: 
   available data, however, do not seem to support this hypothesis since 
   the majority of BALQSO's, once corrected for absorption, have a normal 
   energy distribution. For instance, 
   the 8 BALQSOs observed by Gallagher et al. (\cite{gallagher02}) with 
   {\emph Chandra\/} have a mean optical to 
   X-ray spectral index ${\rm \langle \alpha_{ox} \rangle\,=\,-1.58}$, with 
   an {\it r.m.s.\/} dispersion of 0.11. Similarly, the 6 HI-BAL quasars 
   detected by Green et al. (\cite{green01}) also with {\emph Chandra\/}
   have a mean index ${\rm \langle \alpha_{ox} \rangle\,=\,-1.58}$, with 
   an {\it r.m.s.\/} dispersion of 0.11. Two additional HI-BAL QSO's observed 
   with {\emph XMM-Newton\/} by Grupe et al. (\cite{grupe}) have 
   intrinsic values of 
   ${\rm \alpha_{ox}}$ of -1.50 and -1.48, respectively. All these 
   indices are entirely compatible with the value expected for radio-quiet 
   non-BAL QSOs and much flatter than the upper limit in 
   \object{LBQS 2212-1759}, ${\rm \alpha_{ox} \leq -2.56}$. We note that the
   two LO-BAL quasars detected by Green et al. (\cite{green01}),
   \object{IRAS 07598+6508} and \object{FIRST J0840+3633} have 
   {\em apparent\/}
   steep indices of -2.34 and -2.11, respectively, because they likely suffer 
   from additional absorption which cannot be corrected for given the low
   signal-to-noise ratio of their X-ray spectra. We note that
   \object{LBQS 2212-1759} is definitely not a LO-BAL since its spectrum
   lacks characteristics absorption lines from low ionisation species, 
   such as SiII$\lambda$1808, AlIII$\lambda$1854 and MgII$\lambda$2798
   (see Figure~\ref{abslines}). Only two BALQSO's seem to be 
   genuinely X-ray deficient: 
   with ${\rm \alpha_{ox}\,=\,-1.96}$, \object{PHL~1811} remains marginally 
   X-ray weak after correction for internal absorption, though this conclusion
   is not very robust given the paucity of X-ray photons detected 
   (183: Leighly et al. \cite{leighly01}); the second BALQSO is 
   \object{PG~1254+047} for which Sabra and Hamann (\cite{sabra}) present
   convincing evidence that it is at the same time intrinsically X-ray weak 
   (${\rm \alpha_{ox}\,\leq\,-2.0}$) and heavily absorbed 
   (${\rm N_{H}\,=\,2.8\times 10^{23}\,cm^{2}}$). We note that the indices
   of both sources are still significantly larger than the upper limit
   on ${\rm \alpha_{ox}}$ of \object{LBQS 2212-1759}. One can in principle
   gauge the slope of the optical-to-X-ray spectrum of \object{LBQS 2212-1759}
   from its relative emission line intensities. Indeed, if
   the QSO was lacking EUV and soft X-rays, one would 
   expect the high ionisation emission lines to be relatively weak
   compared to lines from lower ionisation species. In 
   Table~\ref{Lines}, we list the (de-reddened) intensities of the
   main emission lines together with their Full-Width at Half-Maximum (FWHM). 
   The HeII$\lambda$1640 line is undetected, which, at first sight, could
   be indicative of a deficiency in EUV and soft X-rays. We note however, 
   that the upper limit on its intensity relative to that of e.g.
   CIV$\lambda$1249, 0.12, is still higher than that in the average 
   composite SDSS spectrum of Vanden Berk et al. (\cite{vanden berk}), 0.02,
   and therefore not particularly useful.
   Furthermore, the NV$\lambda$1240 line is quite strong, with an intensity 
   relative to Ly$\alpha\lambda$1216, NV/Ly$\alpha$\,$\simeq$\,0.65, 
   compared to e.g. 0.025 in the average composite SDSS spectrum. 
   Since the photon energy required to 
   ionize N$^{3+}$ into N$^{4+}$ is 77.5 eV, much larger than the He$^{+}$ 
   ionisation potential, 24.6 eV, a deficit of EUV and soft X-ray photons 
   seems implausible. In fact, the NIII]$\lambda$1750 is also quite strong 
   (see also Fig.\ref{abslines}), with an intensity relative to
   Ly$\alpha\lambda$1216, III]/Ly$\alpha$\,$\simeq$\,0.12, 32 times larger 
   than that in the average composite SDSS QSO spectrum (Vandeb Berk et al. 
   \cite{vanden berk}). This suggests an overabundance of Nitrogen in
   \object{LBQS 2212-1759}. The CIII]$\lambda$1909/CIV$\lambda$1549 intensity
   ratio, 1.49, is 2.3 times larger than that of the composite SDSS QSO
   (0.63), but this is partly due to absorption eating away  a fraction of 
   the blue wing of the CIV emission line. In summary, the emission line
   intensity ratios do not provide conclusive evidence for a deficit in EUV 
   and soft X-rays in \object{LBQS 2212-1759}.}
   
   
  \item{{\it The LOS to the optical-UV source is different from that to 
   the X-ray source}: with a ``balnicity index'' 
   (Weyman et al. \cite{weyman}) of ${\rm 2221\,km\,s^{-1}}$ 
   (Korista et al. \cite{korista}), \object{LBQS 2212-1759} exhibits  
   relatively weak BAL. Moreover, only about 80~\% of its continuum flux is 
   absorbed at the bottom of the CIV$\lambda$1550 trough, indicating partial 
   coverage of the UV source (Korista et al. \cite{korista}). Similarly, 
   model-fits of the X-ray spectrum of the few BALQSOs detected at high 
   energies also tend to require partial covering of the X-ray source 
   (though models with an ionized absorber yield equally acceptable fits).
   In lower luminosity AGNs, 
   different variability timescales in the two wavebands 
   clearly demonstrate that the X-ray source is $\sim$ an order of magnitude 
   more compact than the UV-optical source. It thus remains a 
   possibility that, in some BALQSOs at least, the outflow  intercepts
   only a fraction of the UV-optical emission while completely blocking-out 
   the X-ray flux. Gallagher et al. (\cite{gallagher04}) reach a similar
   conclusion for the BALQSO \object{PG 2112+059} based on dramatically
   different patterns of variability in the X-ray and the UV regime.
   In \object{PG 2112+059} as well, the absence of a Lyman edge sets an upper
   limit to the UV continuum absorbing column 
   ${\rm N_{H}\,\leq\,10^{17}\,cm^{-2}}$, 5--6 orders of magnitude lower than
   the absorbing column covering the X-ray source.}
   \end{enumerate}

   In the absence of a better choice, hypothesis number two remains our
   favorite explanation for the peculiar SED of \object{LBQS 2212-1759}, 
   though we cannot rule-out the alternative possibility that
   \object{LBQS 2212-1759} is genuinely X-ray weak, with an intrinsically
   steep index, ${\rm \alpha_{ox} \leq -2.56}$. Obviously more observations 
   with {\it Chandra\/} and {\it XMM-Newton\/} are required to increase the 
   number of BALQSOs with measured X-ray properties or at least with 
   stringent upper limits.
  
\begin{acknowledgements}
     The authors are grateful to Kirk Korista for providing his 1991 optical
     spectrum of \object{LBQS 2212-1759} in electronic form. The anonymous
     referee is also thanked for constructive comments which significantly
     improved this article.
     
     This publication makes use of data products from the 
     Two Micron All Sky Survey, which is a joint project of the 
     University of Massachusetts and the Infrared Processing and 
     Analysis Center/California Institute of Technology, funded 
     by the National Aeronautics and Space Administration and the 
     National Science Foundation.
      
     This research has made use of the \emph{NASA/IPAC Extragalactic Database 
     (NED)} which is operated by the Jet Propulsion Laboratory, California 
     Institute of Technology, under contract with the National Aeronautics 
     and Space Administration.
\end{acknowledgements}

\end{document}